\documentclass[preprint,12pt]{elsarticle}

\usepackage{graphicx}
\usepackage{amssymb}
\usepackage{amsmath}
\usepackage{commath}

\usepackage{lineno}
\usepackage{subfigure}
\usepackage{bm}

\usepackage{tikz}
\usetikzlibrary{shapes.geometric, arrows}

\tikzstyle{startstop} = [rectangle, rounded corners, minimum width=3cm, minimum height=1cm,text centered, draw=black, fill=red!30]
\tikzstyle{io} = [trapezium, trapezium left angle=70, trapezium right angle=110, minimum width=3cm, minimum height=1cm, text centered, draw=black, fill=blue!30]
\tikzstyle{process} = [rectangle, minimum width=3cm, minimum height=1cm, text centered, draw=black, fill=orange!30]
\tikzstyle{decision} = [diamond, minimum width=3cm, minimum height=1cm, text centered, draw=black, fill=green!30]
\tikzstyle{arrow} = [thick,->,>=stealth]

\biboptions{sort&compress}

\begin{document}

\begin{frontmatter}


\title{Velocity interpolation based Bounce-Back scheme for non-slip boundary condition in Lattice Boltzmann Method}


\author[uq1]{Pei Zhang}
\author[wu,uk]{S.A. Galindo-Torres}

\author[hhu]{Hongwu Tang\corref{cor1}}
\ead{hwtang@hhu.edu.cn}
\cortext[cor1]{Corresponding author}

\author[hhu]{Guangqiu Jin}
\author[uq1]{A. Scheuermann}
\author[wu]{Ling Li}

\address[uq1]{School of Civil Engineering, University of Queensland, Queensland, Australia\fnref{uq1}}
\address[hhu]{State Key Laboratory of Hydrology-Water Resources and Hydraulic Engineering, Hohai University, Nanjing, China}
\address[uk]{Department of Civil Engineering and Industrial Design, University of Liverpool, Liverpool, UK\fnref{uk}}
\address[wu]{School of Engineering, Westlake University, Hangzhou, China\fnref{wu}}

\begin{abstract}
Lattice Boltzmann Method(LBM) has achieved considerable success on simulating complex flows. However, how to impose correct boundary conditions on the fluid-solid interface with complex geometries is still an open question. Here we proposed a velocity interpolation based bounce-back scheme where the ideas of interpolated bounce-back and non-equilibrium extrapolation are combined. The proposed scheme is validated by several well-defined benchmark cases. It is shown that the proposed scheme offers a better accuracy at high Reynolds number and less dependency on solids positions which may crucial in many engineering and science applications.

\end{abstract}

\end{frontmatter}

%%\linenumbers

\section{Introduction}\label{sec:intro}
Lattice Boltzmann Method(LBM) has emerged as an effective approach of Computational Fluid Dynamics(CFD) during the last decades, and it has attracted numerous interests in simulating complex flows. The success of LBM is mainly contributed by several unique advantages. First, the solution of advection in LBM is exact which reduce the numerical diffusion errors in conventional CFD methods~\cite{van2006galilean}. Second, the locality of collision operator guarantees a high parallelization efficiency of LBM codes. Furthermore, microscopic interactions can be better represented in LBM due to the kinetic nature. However, how to impose correct boundary conditions on the fluid-solid interface (FSI) with complex geometries is still an open question. The most common assumption for FSI is non-slip boundary condition in which the fluid velocities at FSI are equal to solid surface velocities. There are mainly three categories of technologies in LBM to achieve non-slip boundary: Kinetic Boundary Conditions (KBC),  partially saturated cells method (PSM) and Immersed Boundary Method (IBM). 

The simplest KBC is the bounce-back scheme: fluid molecules which contact with the solid surface are reflected back to the fluid domain with opposite velocity. It has been proven that the bounce-back scheme holds the second-order accuracy of LBM in space. However, the bounce-back scheme can only be applied when FSI lies exactly at the nodes or middle of two neighbouring nodes (half-way bounce-back). Otherwise, the real FSI is replaced by an approximated stairwise boundaries which may damage the accuracy of LBM. The idea of using interpolation on distribution functions to reduce geometrical errors is firstly introduced by Bouzidi et al~\cite{bouzidi2001momentum}. This interpolated bounce-back scheme is improved by Yu et al~\cite{yu2003viscous} in which the treatments on distributions are unified regardless of the position of FSI. Filippova~\cite{filippova1998grid} proposed other interpolation scheme and further modified by Mei et al.~\cite{mei1999accurate} to improve the numerical stability. It is found that the relative errors of mentioned interpolation schemes depend on viscosity~\cite{peng2016implementation}, Ginzburg~\cite{ginzburg2003multireflection} developed a multireflection boundary condition which is viscosity-independent. Information at three neighbouring nodes is needed in the multireflection boundary condition. However, there may not always have enough information to implement interpolations in practice, thus Tao et al.~\cite{tao2018one} introduce an one-point second-order curved boundary condition recently. Guo et al.~\cite{guo2002extrapolation} reconstituted the distributions at closest solid nodes by equilibrium and non-equilibrium part, the velocity, density and non-equilibrium are extrapolated from the fluid domain, also known as non-equilibrium extrapolation scheme. All KBCs share some common characters like sharp interfaces, no fluid inside of solids, direct modification on distributions.

The original PSM proposed by Noble et al.~\cite{noble1998lattice,cook2004direct} is designed for particulate flows. The basic idea is mixing the effects of fluid and bounce-back on FSI by volume average. The weighting strategy depends on solid volume fraction where $0$ means fully Saturated and $1$ means fully bounce-back. The biggest advantage of PSM is the smooth transition between fluid and solid nodes, and no refilling algorithms are needed for moving boundaries which is very common for KBCs. Also, the locality and flexibility of PSM are highly desirable for complex flows. For instance, PSM is coupled with Discrete Element Method (DEM)~\cite{feng2007coupled} for dense particulate flows and also for general shaped particles~\cite{galindo2013coupled,galindo2015micro} by using the sphero-polyhedron technique~\cite{galindo2012breaking,galindo2009molecular,galindo2010molecular,galindo2013strength}. Recently, PSM is modified by Zhang et al.~\cite{zhang2017efficient} with Multi-Relaxation model and an efficient particle contact detection strategy.

IBM proposed by Peskin~\cite{peskin2002immersed} also attracted lots of attention due to its flexibility and robustness. Fluid feels solid boundaries by an external force field. Feng et al.~\cite{feng2004immersed} combined IBM with LBM firstly, the penalty method is employed to link flows and particle motions. The IBM is enhanced by Luo et al.~\cite{luo2007modified} where velocity distributions at the boundary layer are introduced to improve the accuracy near particle surfaces. The idea of momentum exchange is also combined with IBM to avoid user-defined parameters in penalty methods~\cite{niu2006momentum}. Wu et al.~\cite{wu2009implicit} notice that the non-slip condition is not exactly satisfied in explicit IBMs, thus an implicit velocity correction based IBM is proposed. Most of IBMs have a diffuse interface due to the smoothed external force field, the sharp interface can also be achieved in IBM as shown in~\cite{kang2011comparative}. 

As mentioned, the sharp interface can only be found in KBCs and servral IBMs, thus KBCs are surposed to be more accurate. Here we presented an improved version of KBC where the ideas of interpolated bounce-back schemes and non-equilibrium extrapolation scheme are combined, the proposed scheme shows better accuracy at high Reynolds number and less dependency on solids positions which may crucial in many appilcations.

The structure of the paper is organized as follows: Sec.~\ref{sec:lbm} describes the basics of LBM. Sec.~\ref{sec:bc} explains the ideas and approximations of proposed scheme. And validations are presented in Sec.~\ref{sec:benchmark} with several well-defined benchmark cases. Finally Sec.~\ref{sec:conclusion} presents conclusions from the present work.

\section{Lattice Boltzmann method for fluid}\label{sec:lbm}
Flows are solved by the Lattice Boltzmann Method~\cite{galindo2012numerical,galindo2013lattice} with the D2Q9 model where spaces are divided into square lattices and the velocity domain is discretized into 9 discrete velocity vectors as follows: 

\begin{equation*}
\resizebox{0.5\hsize}{!}
{$\overrightarrow{e}_i = \left\{ 
  \begin{array}{l l l}
    0, & \quad \text{$i$ = 0,}\\[0.5ex]
    (\pm1,0,),(0,\pm1), & \quad \text{$i$ = 1 to 4,}\\[0.5ex]
    (\pm1,\pm1), & \quad \text{$i$ = 5 to 8,}\\
  \end{array} \right.$}
\end{equation*}

Based on the Chapman-Enskog expansion of the Boltzmann equation, an evolution rule is applied to every distribution function~\cite{mohamad2011lattice}:
\begin{equation}
f_i(\overrightarrow{x}+\overrightarrow{e}_i\delta{t},t+\delta{t}) = f_i(\overrightarrow{x},t) + \Omega_{col},
\end{equation}
where $f_i$ is the probability distribution function, $\overrightarrow{x}$ is the position of the local lattice, $\delta{t}$ is the time step and $\Omega_{col}$ is the collision operator. The most widely used form of $\Omega_{col}$ is the so-called Bhatnagar-Gross-Krook (BGK) collision operator: $\Omega_{col} = \frac{\delta{t}}{\tau}(f^{eq}_{i}-f_i)$, with $f^{eq}_{i}$ the equilibrium distribution given by,
\begin{equation}
f^{eq}_{i} = \omega_i\rho \bigg(1 + 3\frac{\overrightarrow{e}_i \cdot \overrightarrow{u}}{C^2} + \frac{9(\overrightarrow{e}_i \cdot \overrightarrow{u})^2}{2C^4} - \frac{3u^2}{2C^2}\bigg),
\label{eq:feq}
\end{equation}
where $C=\delta{x}/\delta{t}$ is the characteristic lattice velocity ($\delta{x}$ is the lattice size), here we choose $C=\delta{x}=\delta{t}=1$. The weights are $\omega_0=4/9$, $\omega_i=1/9$ for $i=$1 to 4, and $\omega_i=1/36$ for $i=$5 to 8.

The BGK collision operator assumes that the collision only depends on a dimensionless relaxation time $\tau$, where $\tau=3\nu+0.5$. However, It's found that the simulations become unstable when the value of $\tau$ is close to 0.5~\cite{galindo2013coupled}. Therefore, the BGK collision operator is only suitable for flow at relatively low Reynolds numbers. To overcome this limitation, the multiple relaxation time (MRT) collision operator is adopted in this study as follows:

\begin{equation}
\Omega_{col} = \mathbf{\hat{M}^{-1}\hat{S}}(m^{eq}_{i}-m_i),
\label{eq:mrt}
\end{equation}
with $m_i=\mathbf{\hat{M}}f_i$, where $\mathbf{\hat{M}}$ is a matrix used to transform the probability distribution function $f_i$ to velocity moments linearly. For the D2Q9 model, the moments are arranged as: $m_0=\rho$; $m_1=e$; $m_2=\epsilon$; $m_{3,5}=j_{x,y}$ are components of the momentum $\overrightarrow{j}=(j_x,j_y)=\rho \overrightarrow{u}$; $m_{4,6}=q_{x,y}$ are related to components of the heat flux $\overrightarrow{q}=(q_x,q_y)$; $m_7=p_{xx}$; and $m_8=p_{xy}$ are related to the components of the strain-rate tensor.  The equilibrium moments are the functions of conserved moments (density $\rho$ and moment density $\overrightarrow{j}$) and the non-conserved moments are given by~\cite{lallemand2000theory},
\begin{equation}
\begin{array}{cc}
m^{eq}_1=e^{eq}=\rho(-2+3\overrightarrow{j}\cdot\overrightarrow{j}), & m^{eq}_2=\epsilon^{eq}=\rho(1-3\overrightarrow{j}\cdot\overrightarrow{j}), \\[2ex]
m^{eq}_4=q_x^{eq}=-j_x, & m^{eq}_6=q_y^{eq}=-j_y, \\[2ex]
m^{eq}_7=p_{xx}^{eq}=\frac{{j_x}^2-{j_y}^2}{\rho}, & m^{eq}_8=p_{xy}^{eq}=\frac{j_x j_y}{\rho},
\end{array}
\end{equation}
the transformation matrix is defined as:
\begin{equation}
\mathbf{\hat{M}} = \begin{bmatrix}
       \phantom{-}1 & \phantom{-}1 & \phantom{-}1 & \phantom{-}1 & \phantom{-}1 & \phantom{-}1 & \phantom{-}1 & \phantom{-}1 & \phantom{-}1          \\[0.3em]
       -4&-1 &-1 &-1 &-1 & \phantom{-}2 & \phantom{-}2 & \phantom{-}2 & \phantom{-}2          \\[0.3em]
       \phantom{-}4 &-2 &-2 &-2 &-2 & \phantom{-}1 & \phantom{-}1 & \phantom{-}1 & \phantom{-}1          \\[0.3em]
       \phantom{-}0 & \phantom{-}1 & \phantom{-}0 &-1 & \phantom{-}0 & \phantom{-}1 &-1 &-1 & \phantom{-}1          \\[0.3em]
       \phantom{-}0 &-2 & 0 & \phantom{-}2 & \phantom{-}0 & \phantom{-}1 &-1 &-1 & \phantom{-}1          \\[0.3em]
       \phantom{-}0 & \phantom{-}0 & \phantom{-}1 & \phantom{-}0 &-1 & \phantom{-}1 & \phantom{-}1 &-1 & \phantom{-}1          \\[0.3em]
       \phantom{-}0 & \phantom{-}0 &-2 & \phantom{-}0 & \phantom{-}2 & \phantom{-}1 & \phantom{-}1 &-1 &-1          \\[0.3em]
       \phantom{-}0 & \phantom{-}1 &-1 & \phantom{-}1 &-1 & \phantom{-}0 & \phantom{-}0 & \phantom{-}0 & \phantom{-}0          \\[0.3em]
       \phantom{-}0 & \phantom{-}0 & \phantom{-}0 & \phantom{-}0 & \phantom{-}0 & \phantom{-}1 &-1 & \phantom{-}1 &-1          
     \end{bmatrix}
     \label{eq:M}
\end{equation}

In Eq.~\ref{eq:mrt}, $\mathbf{\hat{S}}$ is the diagonal relaxation matrix in velocity moments. The kinetic viscosity is related to $\mathbf{\hat{S}}$, the diagonal elements of $\mathbf{\hat{S}}$ is given as:
\begin{equation*}
\resizebox{0.45\hsize}{!}
{$s_{i,i} = \left\{ 
  \begin{array}{l l l}
    0.3, & \quad \text{$i$ = 0,3,5}\\[0.5ex]
    1.5, & \quad \text{$i$ = 1,2}\\[0.5ex]
    1.2, & \quad \text{$i$ = 4,6}\\[0.5ex]
    \frac{1}{3\nu+0.5}, & \quad \text{$i$ = 7,8}\\
  \end{array} \right.$}
\end{equation*}

Here the Mach number is defined as the ratio of the maximum velocity to $C$. When $Ma\ll1$, the LBE can be used to recover the Navier-Stokes equation. More detail can be found in~\cite{mohamad2011lattice}. The macroscopic fluid properties such as density $\rho$ and flow velocity $\overrightarrow{u}$ can be determined by the zero-th and the first order moment of the distribution function: 
\begin{equation}
\begin{array}{ll}
\rho(\overrightarrow{x}) &= \sum_{i=0}^{8} f_i(\overrightarrow{x}),\\[2ex]
\overrightarrow{u}(\overrightarrow{x}) &= \frac{1}{\rho(\overrightarrow{x})} \sum_{i=0}^{8} f_i(\overrightarrow{x})\overrightarrow{e}_i ,
\end{array}
\end{equation}

\section{fluid-solid interface boundary conditions}\label{sec:bc}
\begin{figure}[t]
\begin{centering}
\includegraphics[width=0.8\linewidth]{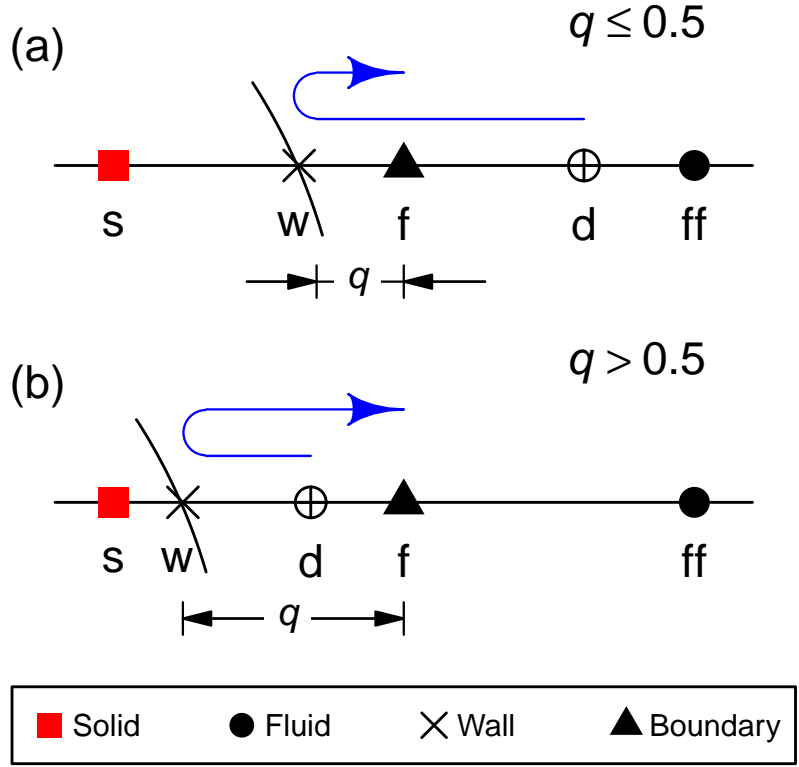}
\caption{Schematic of the bounce back role at FSI, where ``s" for the closest solid node, ``w" for wall, ``f" for the boundary node, ``ff" for the neighbouring fluid node of ``f", ``d" for the depart position where molecules arrive to ``f" at next time step.}
\label{fig:ibb}
\end{centering}
\end{figure}
The no-penetration non-slip boundary condition needs to be imposed on the fluid-solid interface properly. There are mainly two categories of boundary schemes can satisfy the requirements. Firstly, macroscopic boundary conditions where macroscopic properties are modified, such as velocity corrected immersed boundary method in which the effects of boundaries are replaced by a smoothed external force field for fluid. In the second category, the distribution functions are used directly to impose boundary conditions. Here we consider the later one since it can maintain a sharp interface and fit the kinetic nature of LBM.

The computational nodes are divided into fluid nodes and solid nodes, the fluid nodes which are close to the solid boundary are further identified as boundary nodes (Fig.~\ref{fig:ibb}). Since the uniform-sized mesh is used in classic LBM, the curved boundaries generally locate between boundary nodes and solid nodes. Thus the distribution functions at boundary nodes which streamed from solid nodes are missing, the key task is to determine missing distribution functions properly.

The simplest solution is the bounce-back role where molecules depart from $f$ with velocity $\bm{e}_{i'}$ hit on wall and return back to node $f$ with opposite discrete velocity ($\bm{e}_{i}$). It is clear that the wall is assumed to be located at the middle point between point $s$ and $f$ regardless of the actual position. This assumption leads to stairwise boundaries which damage the second order accuracy of LBM. Therefore, Bouzidi et al~\cite{bouzidi2001momentum} proposed an interpolated bounce-back scheme to reduce geometrical errors. As shown in Fig.~\ref{fig:ibb}, molecules depart from $d$ with velocity $\bm{e}_{i'}$ will end up to node $f$ after bounce-back on the wall, the position of $d$ depends on $q=\abs{\bm{x}_f-\bm{x}_w}/\abs{\bm{x}_f-\bm{x}_s}$. For $q \leqslant 0.5$, the distribution function after collision $f^{+}_{i'}(\bm{x}_d,t)$ can be linearly interpolated using the information at node $f$ and $ff$. For $q>0.5$, $d$ lies between $f$ and $w$ where no information of distributions are known. Also, the exploration can be used but it's unfavourable due to numerical stability issues and unbounded errors. After steaming, molecules depart from $f$ with velocity $\bm{e}_{i'}$ will end up to $d$, thus $f_{i}(\bm{x}_f,t+\delta{t})$ can be determined based on $f_{i}(\bm{x}_d,t+\delta{t})$ and $f_{i}(\bm{x}_ff,t+\delta{t})$. Bouzidi's scheme can be summarized as:

\begin{equation}
f_{i}(\bm{x}_f,t+\delta{t}) = 2qf^{+}_{i'}(\bm{x}_f,t) + (1-2q)f^{+}_{i'}(\bm{x}_{ff},t) + 6\omega_{i'}\rho_{0}\frac{\bm{e}_i \cdot \bm{u}_{w}}{C^2}, \quad \text{$q \leqslant 0.5$}
\label{eq:q_less}
\end{equation}

\begin{equation}
f_{i}(\bm{x}_f,t+\delta{t}) = \frac{1}{2q}(f^{+}_{i'}(\bm{x}_f,t)+ 6\omega_{i'}\rho_{0}\frac{\bm{e}_i \cdot \bm{u}_{w}}{C^2}) + \frac{2q-1}{2q}f_{i}(\bm{x}_{f},t), \quad \text{$q > 0.5$}
\label{eq:q_larger}
\end{equation}
%\begin{equation*}
%\resizebox{1.\hsize}{!}
%{$f_{i}(\bm{x}_f,t+\delta{t}) = \left\{ 
%  \begin{array}{l l}
%    \text{$2qf^{+}_{i'}(\bm{x}_f,t) + (1-2q)f^{+}_{i'}(\bm{x}_{ff},t) + 6\omega_{i'}\rho_{0}\frac{\bm{e}_i \cdot \bm{u}_{w}}{C^2}$}, & \quad \text{$q \leqslant 0.5$,}\\[3ex]
%     \text{$\frac{1}{2q}(f^{+}_{i'}(\bm{x}_f,t)+ 6\omega_{i'}\rho_{0}\frac{\bm{e}_i \cdot \bm{u}_{w}}{C^2}) + \frac{2q-1}{2q}f_{i}(\bm{x}_{f},t)$}, & \quad \text{$q > 0.5$,}\\[0.ex]
%  \end{array} \right.$}
%  \label{eq:bouzidi}
%\end{equation*}
where $\bm{u}_w$ is the wall velocity, notice that the term in Eq.~\ref{eq:q_less} and~\ref{eq:q_larger} which including $\bm{u}_w$ indicates the momentum exchange due to the moving wall~\cite{ladd1994numerical}.

Yu et al~\cite{yu2003viscous} proposed an unified interpolated bounce-back scheme regardless of the value of $q$. The idea is to evaluate distributions at wall $f_{i'}(\bm{x}_w,t+\delta{t})$ first, then the bounce-back role is applied, the missing distributions at $f$ after streaming $f_{i}(\bm{x}_f,t+\delta{t})$ is interpolated between $f_{i}(\bm{x}_w,t+\delta{t})$ and $f_{i}(\bm{x}_ff,t+\delta{t})$. Yu's scheme can be summarized as:
\begin{equation}
f_{i}(\bm{x}_w,t+\delta{t}) = qf^{+}_{i'}(\bm{x}_f,t) + (1-q)f^{+}_{i'}(\bm{x}_{ff},t) + 6\omega_{i'}\rho_{0}\frac{\bm{e}_i \cdot \bm{u}_{w}}{C^2},
\end{equation}

\begin{equation}
f_{i}(\bm{x}_f,t+\delta{t}) = \frac{1}{1+q}(f_{i}(\bm{x}_w,t+\delta{t}) + \frac{q}{1+q}f_{i}(\bm{x}_{ff},t+\delta{t}),
\end{equation}

\subsection{velocity interpolation based bounce-back scheme}\label{sec:vibb}
Beside of interpolated bounce-back schemes, Guo et al~\cite{guo2002extrapolation} introduced a non-equilibrium extrapolation boundary condition, where virtual distributions at solid node $s$ are decomposed into equilibrium part $f^{eq}_{i}(\bm{x}_s,t)$ and non-equilibrium part $f^{neq}_{i}(\bm{x}_s,t)$. $f^{eq}_{i}(\bm{x}_s,t)$ can be determined by Eq.~\ref{eq:feq} with $\bm{u}_s$ and $\rho_s$. Both $\bm{u}_s$ and $f^{neq}_{i}(\bm{x}_s,t)$ are extrapolated from fluid nodes $f$ and $ff$. Since the fluctuation of density is of order $O({Ma}^2)$, $\rho_s$ is approximated as $\rho_s=\rho_f$. The main idea behind Guo's scheme is the fact that the distributions are dominated by the equilibrium part since the variations of $f^{neq}$ are one order smaller than $f^{eq}$. Thus it is safe to extrapolate $f^{neq}$ with second order accuracy~\cite{guo2002extrapolation}.

Here we combine the idea of interpolated bounce-back and non-equilibrium extrapolation/interpolation together. As shown in Fig.~\ref{fig:ibb}, the unkown $f_{i}(\bm{x}_f,t+\delta{t})$ is determined by bounce-back role:
\begin{equation}
f_{i}(\bm{x}_f,t+\delta{t}) = f^{+}_{i'}(\bm{x}_d,t) + 6\omega_{i'}\rho_{0}\frac{\bm{e}_i \cdot \bm{u}_{w}}{C^2},
\label{eq:eqneq}
\end{equation}
where $f^{+}_{i'}(\bm{x}_d,t)$ is given as:
\begin{equation}
f^{+}_{i'}(\bm{x}_d,t) = f^{eq}_{i'}(\rho_d,\bm{u}_d) + f^{neq}_{i'}(\bm{x}_d,t),
\label{eq:eqneq}
\end{equation}
$\rho_d$ is extrapolated/interpolated as
\begin{equation}
\rho_d = 2q\rho_f + (1-2q)\rho_{ff},
\label{eq:rhod}
\end{equation}
And $f^{neq}_{i'}(\bm{x}_d,t)$ is handled in the same way and rescaled by density ratio:
\begin{equation}
f^{neq}_{i'}(\bm{x}_d,t) = 2q(f^{+}_{i'}(\bm{x}_f,t)-f^{eq}_{i'}(\bm{x}_f,t))\frac{\rho_d}{\rho_f} + (1-2q)(f^{+}_{i'}(\bm{x}_{ff},t)-f^{eq}_{i'}(\bm{x}_{ff},t))\frac{\rho_{d}}{\rho_{ff}},
\label{eq:neqd}
\end{equation}

\begin{figure}[t]
\begin{centering}
\includegraphics[width=0.8\linewidth]{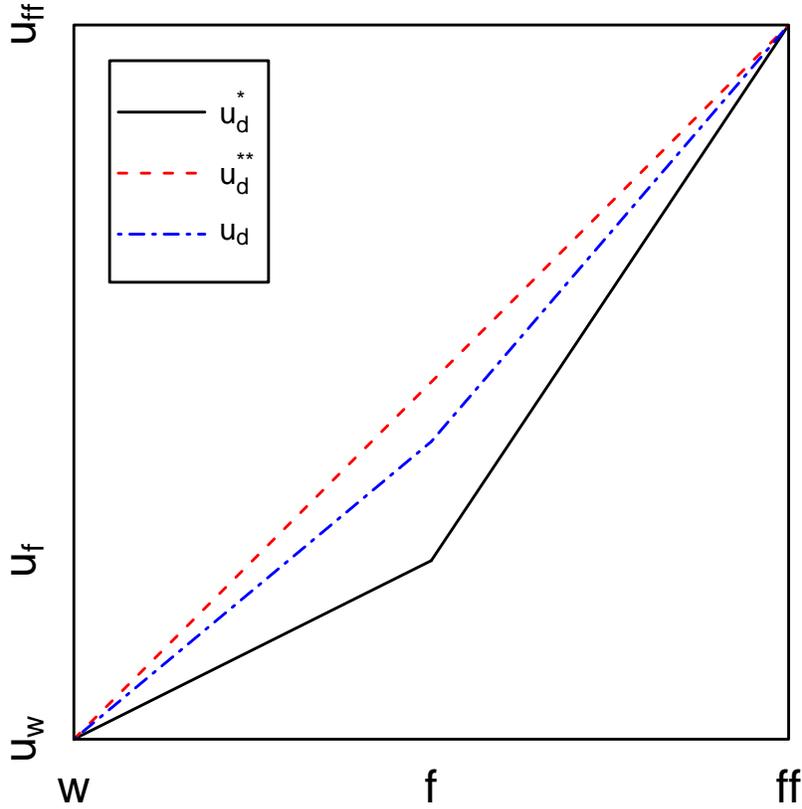}
\caption{Schematic of velocity interpolation schemes.}
\label{fig:interpolation}
\end{centering}
\end{figure}

As shown in Guo's scheme, velocities play the most important roles in determining unknown distributions. Fortunately, both $\bm{u}_w$, $\bm{u}_f$ and $\bm{u}_{ff}$ are known. $\bm{u}_d$ in Eq.~\ref{eq:eqneq} can be evaluated by linear interpolation separately:
\begin{equation}
\resizebox{0.6\hsize}{!}
{$\bm{u}^{*}_d = \left\{ 
  \begin{array}{l l}
    \text{$2q\bm{u}_f + (1-2q)\bm{u}_{ff}$}, & \quad \text{$q \leqslant 0.5$,}\\[3ex]
     \text{$\frac{1-q}{q}\bm{u}_f + \frac{2q-1}{q}\bm{u}_w$}, & \quad \text{$q > 0.5$,}\\[0.ex]
  \end{array} \right.$}
  \label{eq:bouzidi}
\end{equation}
or linearly interpolated between $\bm{u}_w$ and $\bm{u}_{ff}$ regardless of $\bm{u}_{f}$:
\begin{equation}
\bm{u}^{**}_d = \frac{1-q}{1+q}\bm{u}_{ff} + \frac{2q}{1+q}\bm{u}_w,
\label{eq:neqd}
\end{equation}
$\bm{u}_d$ is calculated by weighted averange as shown in Fig.~\ref{fig:interpolation}:
\begin{equation}
\bm{u}_d = \frac{1}{3}\bm{u}^{*}_d + \frac{2}{3}\bm{u}^{**}_d,
\label{eq:neqd}
\end{equation}

We choose above interpolation scheme because of the following observation: $\bm{u}_w$ and $\bm{u}_{ff}$ are supposed to be more accurate since they are not affected by the unknown distributions. Thus $\bm{u}^{*}_d$ which including $\bm{u}_{f}$ is assigned with less weight. Notes that above boundary scheme cannot be recovered to the bounce-back scheme when $q=0.5$. However, Peng et al~\cite{peng2016implementation} report that errors of linear interpolated bounce-back increase with $q$, thus recovering to bounce-back scheme do not guarantee a better accuracy. Later, we will show this inconsistency have trivial effects on results.

\section{Benchmark and discussion}\label{sec:benchmark}
\subsection{Poiseuille flow with a moving wall}\label{sec:poiseuille}
\begin{figure}[t]
\begin{centering}
\includegraphics[width=0.8\linewidth]{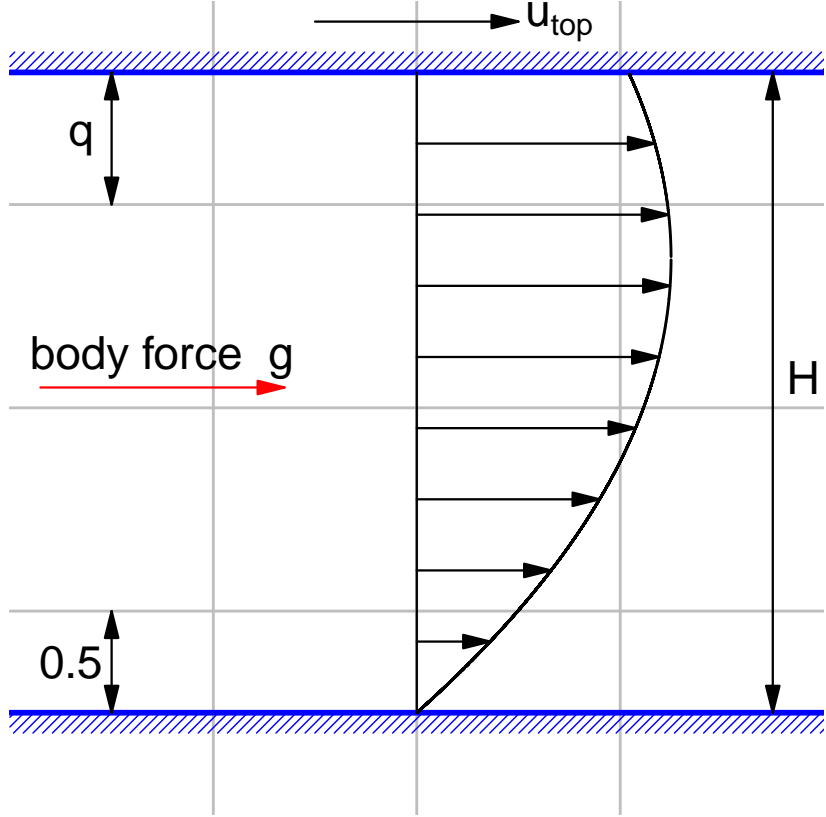}
\caption{Schematic of Poiseuille flow with a moving wall.}
\label{fig:poiseuille}
\end{centering}
\end{figure}

To evaluate the accuracy of the proposed boundary condition with well-defined geometries, the Poiseuille flow with a moving wall is chosen as the first benchmark case as shown in Fig.~\ref{fig:poiseuille}. The bottom wall is fixed and halfway bounce-back is applied, while the top wall is moving with $u_{top}$ and different boundary conditions are tested. The fluid is initially rest and driven by a constant body force $g$ along the x-direction. The analytical solution for steady state is given as\citep{peng2016implementation}
\begin{equation}
u_x = -\frac{g}{2\nu}y^2+\bigg(\frac{g}{2\nu}H + \frac{u_{top}}{H}\bigg)y,
\label{eq:exact}
\end{equation}
where $H$ is the height of the channel and $y$ is the vertical coordinate. A $L \times L$ domain size is used and the top wall is placed above the highest lattices with distance $q$. $q$ varies between 0 and 1. Since $H$ varies with $q$, the Reynolds number is defined as $Re=u_{top}L/\nu$. The parameters are chosen as: $L=60$, $g=1.0\times 10^{-6}$, $u_{top}=0.01$. Two values of viscosity are used $\nu=0.1$ and $\nu=0.01$ which correspond $Re=6$ and $60$.

\begin{figure}[t]
\begin{centering}
\includegraphics[width=0.8\linewidth]{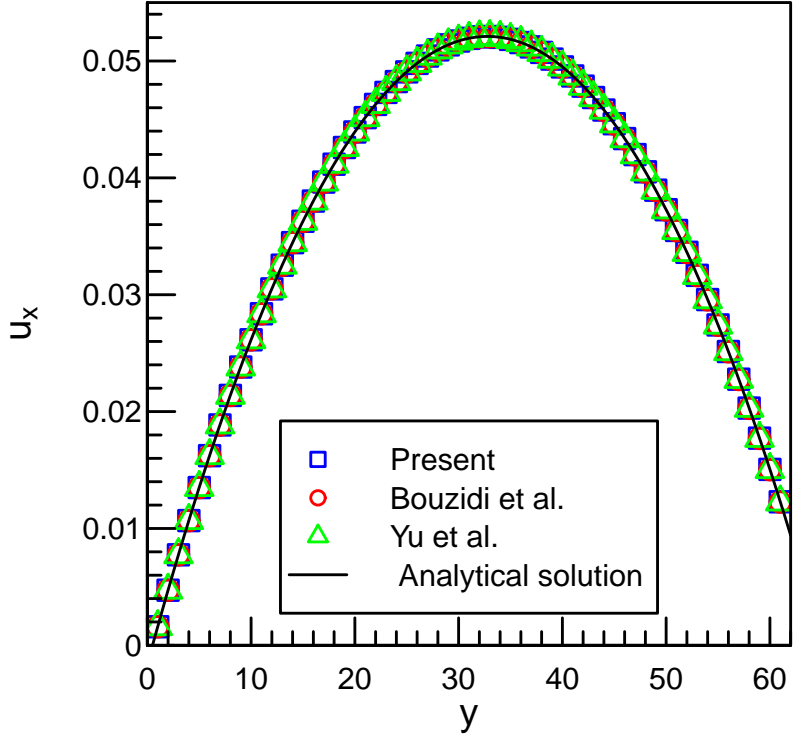}
\caption{Simulated velocity profile at $Re=60$ and $q=0.8$.}
\label{fig:vel_re60}
\end{centering}
\end{figure}

\begin{figure}[t]
\begin{centering}
\includegraphics[width=0.8\linewidth]{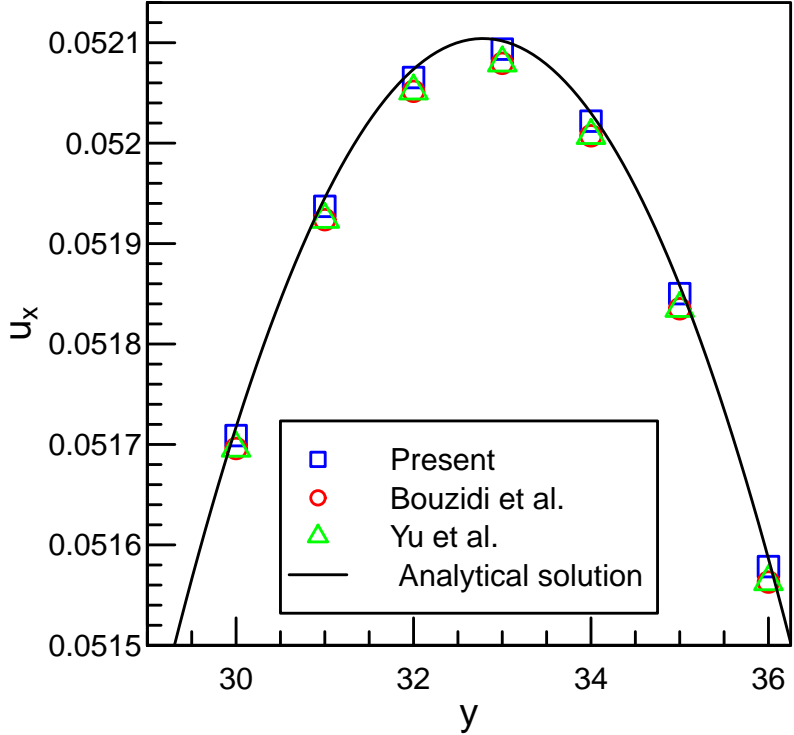}
\caption{A zoom-in view of Fig.~\ref{fig:vel_re60}.}
\label{fig:vel_zoom_re60}
\end{centering}
\end{figure}

\begin{figure}[t]
\begin{centering}
\includegraphics[width=0.8\linewidth]{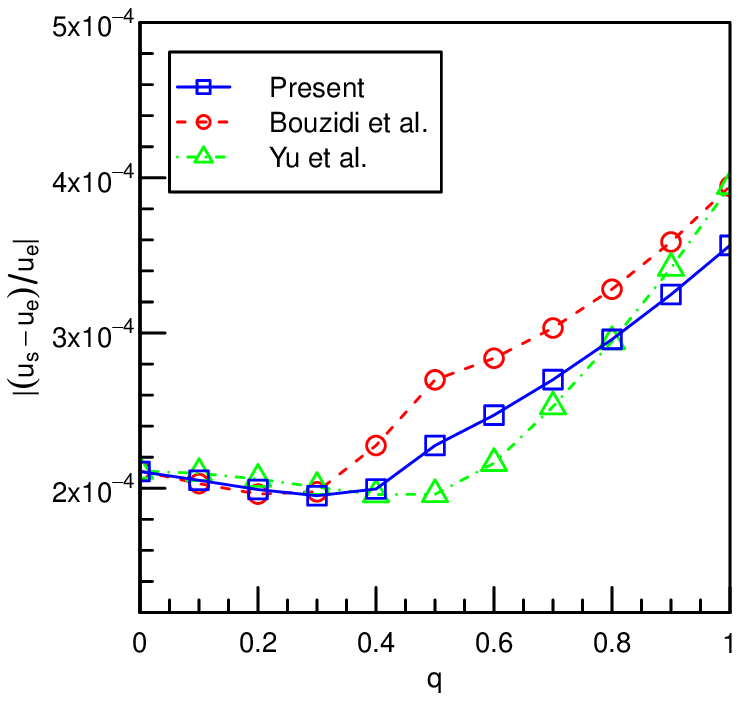}
\caption{Errors as a function of $q$ at $Re=6$.}
\label{fig:error_re6}
\end{centering}
\end{figure}

\begin{figure}[t]
\begin{centering}
\includegraphics[width=0.8\linewidth]{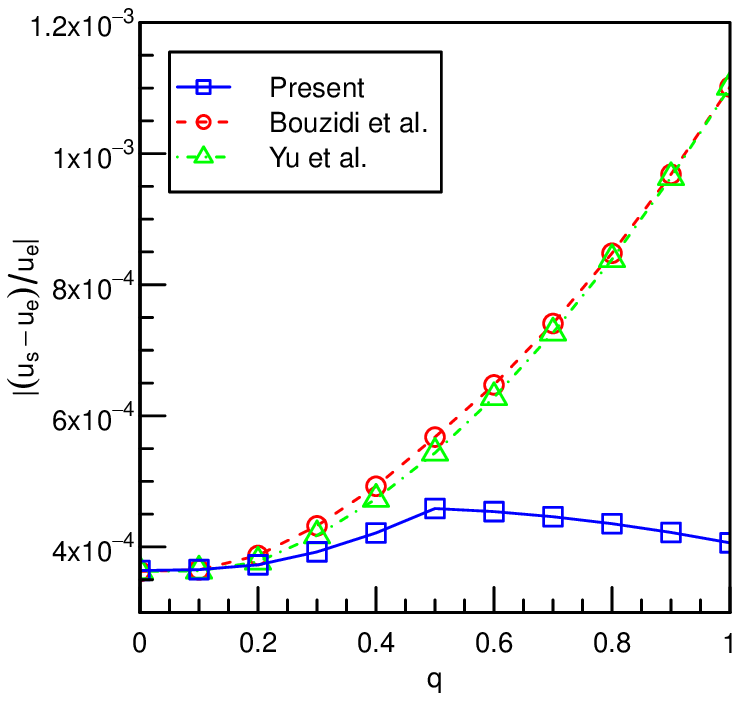}
\caption{Errors as a function of $q$ at $Re=60$.}
\label{fig:error_re60}
\end{centering}
\end{figure}

\begin{figure}[t]
\begin{centering}
\includegraphics[width=0.8\linewidth]{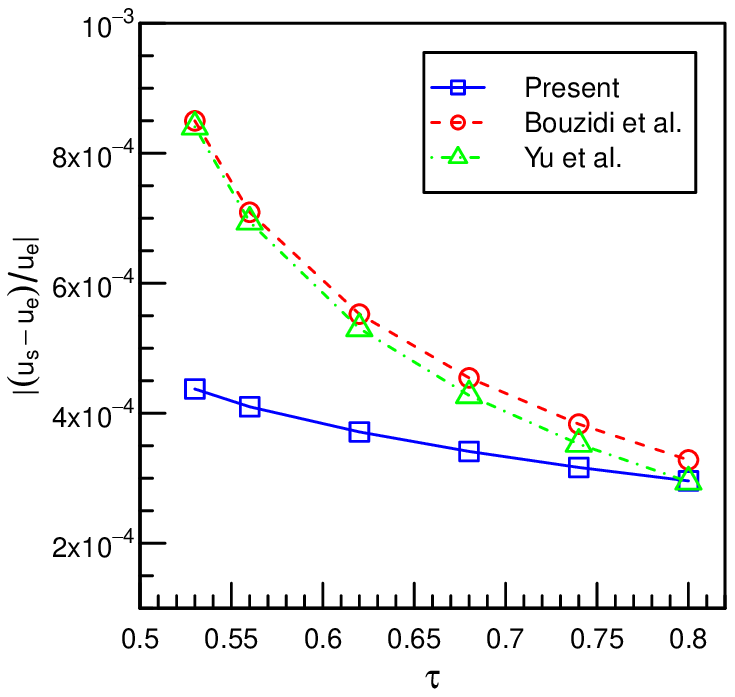}
\caption{$\tau$ effect on errors at $q=0.8$.}
\label{fig:error_tau}
\end{centering}
\end{figure}

Simulated velocity profile are compared with the analytical solution in Fig~\ref{fig:vel_re60} at $Re=60$ and $q=0.8$. Overall, all boundary schemes provide reasonable results. However, the proposed scheme shows a better accuracy compared with Bouzidi and Yu's scheme in the zoom-in view (Fig~\ref{fig:vel_zoom_re60}). Previous studies~\cite{peng2016implementation} indicate that errors depend on the value of $q$, Fig.~\ref{fig:error_re6} and Fig.~\ref{fig:error_re60} show errors vary with $q$ at $Re=6$ and $Re=60$. Here the error is defined as:$\left| (u_s - u_e)/u_e \right|$ where $u_s$ is simulated velocity and $u_e$ is the exact value from Eq.~\ref{eq:exact}. At low $Re$, all three schemes show identical pattern at small $q$ while the proposed scheme shows slightly less error at large $q$. In practice, the global error mainly depends on the maximum local error, thus a small improvement on maximum local error can still considerably increase the overall accuracy. At high $Re$, it is clear that the proposed scheme performs much better, especially at large $q$. Furthermore, errors decrease with increasing $q$ when $q>0.5$ and almost one order magnitude smaller than other schemes. Peng et al~\cite{peng2016implementation} argue that quadratic interpolation schemes have better consistency than linear schemes since quadratic schemes have converging error at $q=0$ and $q=1$. The proposed scheme also shows significant improvements in terms of consistency at high $Re$. These improvements are mainly due to the usage of $\bm{u}_w$ which eliminate the unphysical slipping on boundaries. To investigate the effect of $\tau$, $q$ is fixed to $0.8$ and $\tau$ varies between $0.5$ and $1$ which is the typical range of $\tau$ in practices. As shown in Fig.~\ref{fig:error_tau}, the proposed scheme shows less dependence on $\tau$ than others.

\subsection{Cylindrical Couette flow}\label{sec:couette}
\begin{figure}[t]
\begin{centering}
\includegraphics[width=0.8\linewidth]{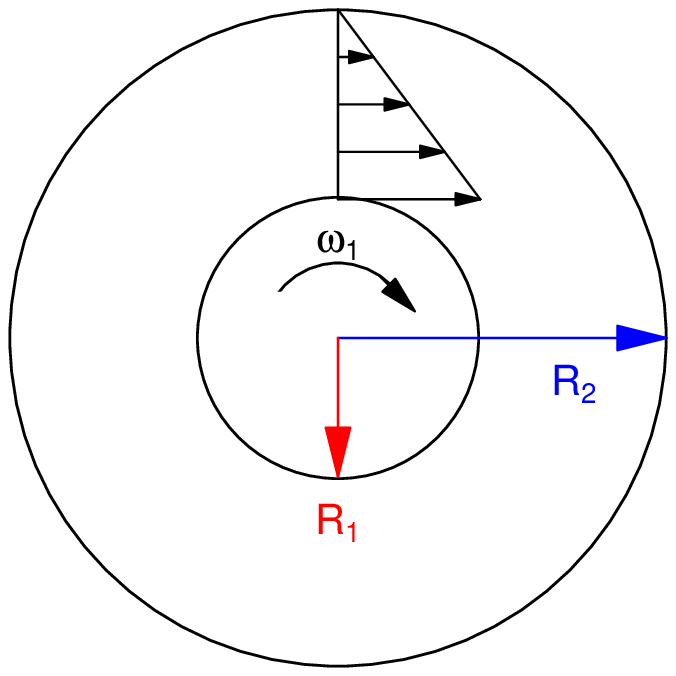}
\caption{Schematic of cylindrical Couette flow.}
\label{fig:couette}
\end{centering}
\end{figure}

Another classic benchmark is the cylindrical Couette flow which involves curved boundaries as shown in Fig.~\ref{fig:couette}. Two cylinders are placed at the centre of the domain ($L \times L$), where the inner cylinder rotating with a constant angular velocity $\omega_1$ and the outer cylinder is fixed. To evaluate the order of accuracy of the proposed scheme, the radius of cylinders are set as $R_1=L/4.8$ and $R_2=L/2.4$. The analytical solution for steady state is given as
\begin{equation}
\frac{u_r}{\omega_1 R_1} = \frac{R_1 R_2}{R_2^2-R_1^2}\bigg(\frac{R_2}{r}-\frac{r}{R_2}\bigg),
\label{eq:exact2}
\end{equation}
where $u_r$ is the velocity component which perpendicular to the radial direction and $r$ is the radial distance from the centre of cylinders. Four values of $L$ are used in simulations: $20, 40, 80, 160$ and viscosity varies with $\omega_1$ to fix Reynolds number as $Re=6$ and $60$.

\begin{figure}[t]
\begin{centering}
\includegraphics[width=0.8\linewidth]{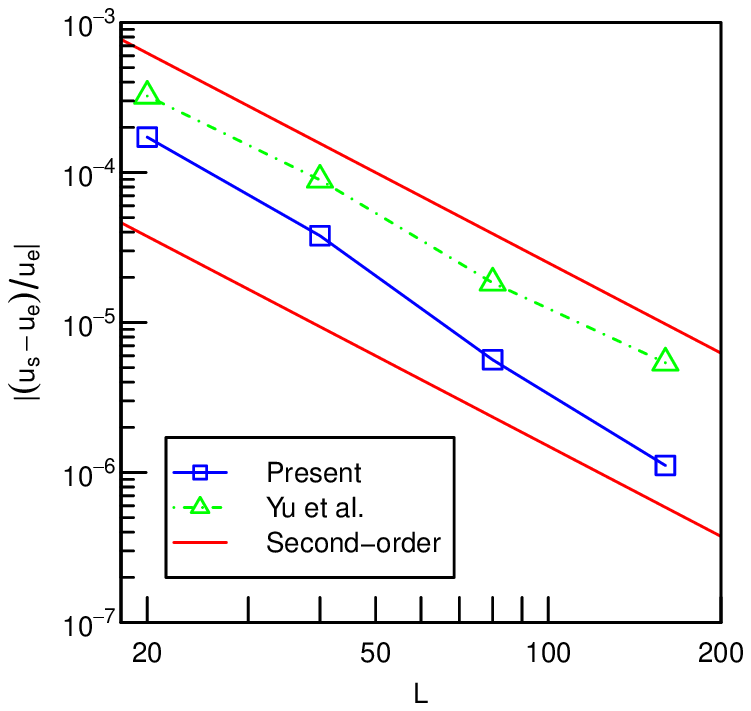}
\caption{Errors as a function of resolutions at $Re=6$.}
\label{fig:error_vs_l_re6}
\end{centering}
\end{figure}

\begin{figure}[t]
\begin{centering}
\includegraphics[width=0.8\linewidth]{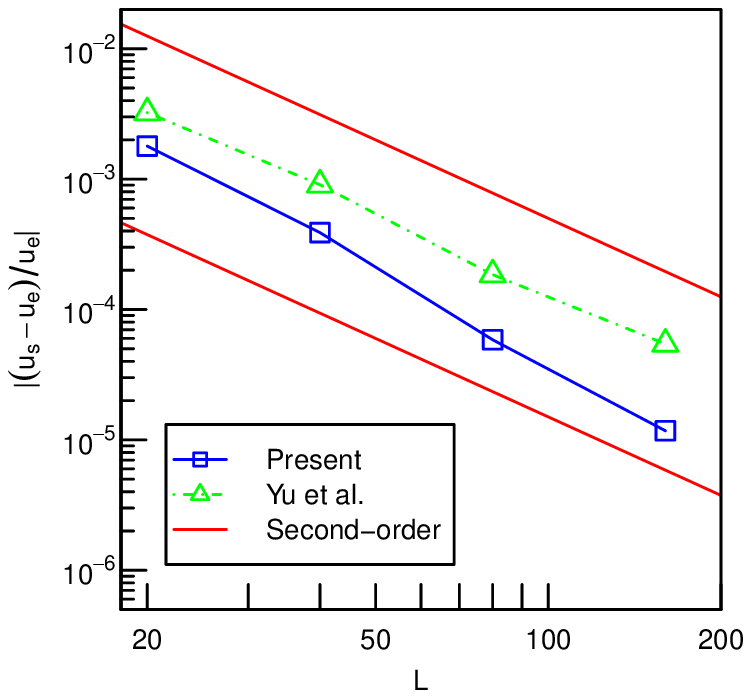}
\caption{Errors as a function of resolutions at $Re=60$.}
\label{fig:error_vs_l_re60}
\end{centering}
\end{figure}

\begin{figure}[t]
\begin{centering}
\includegraphics[width=0.8\linewidth]{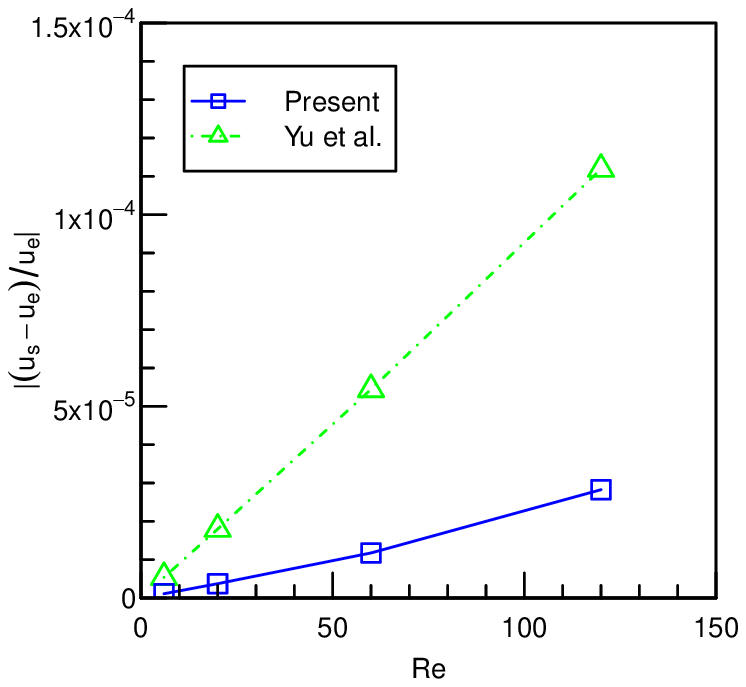}
\caption{Errors as a function of Reynolds number.}
\label{fig:error_vs_re}
\end{centering}
\end{figure}

The relative errors against resolutions are plotted on a log scale in Fig.~\ref{fig:error_vs_l_re6} and Fig.~\ref{fig:error_vs_l_re60} for $Re=6$ and $Re=60$ respectively. The results of Yu's scheme are also presented as a comparison. The results confirm that interpolated bounce-back schemes are generally second-order accuracy. The proposed scheme shows more accurate at all resolutions. Compared to Fig.~\ref{fig:error_re6} in which all schemes have identical errors at low $Re$, it is surprising that the proposed scheme performs better regardless of $Re$. This superiority can be explained by the fact that errors increase with $q$ thus little improvements for large $q$ can significantly increase accuracy. It can be also found in Fig.~\ref{fig:error_re6} and~\ref{fig:error_re60} that the accuracy of the proposed scheme is slightly better than second-order, especially at high resolution. The errors are found linearly increased with $Re$ as shown in Fig.~\ref{fig:error_vs_re} ($L=160$), but the proposed scheme shows less dependence on $Re$ where errors of Yu's scheme increases dramatically at high $Re$ in comparison. 

\subsection{Magnus effect}\label{sec:magnus}
\begin{figure}[t]
\begin{centering}
\includegraphics[width=0.8\linewidth]{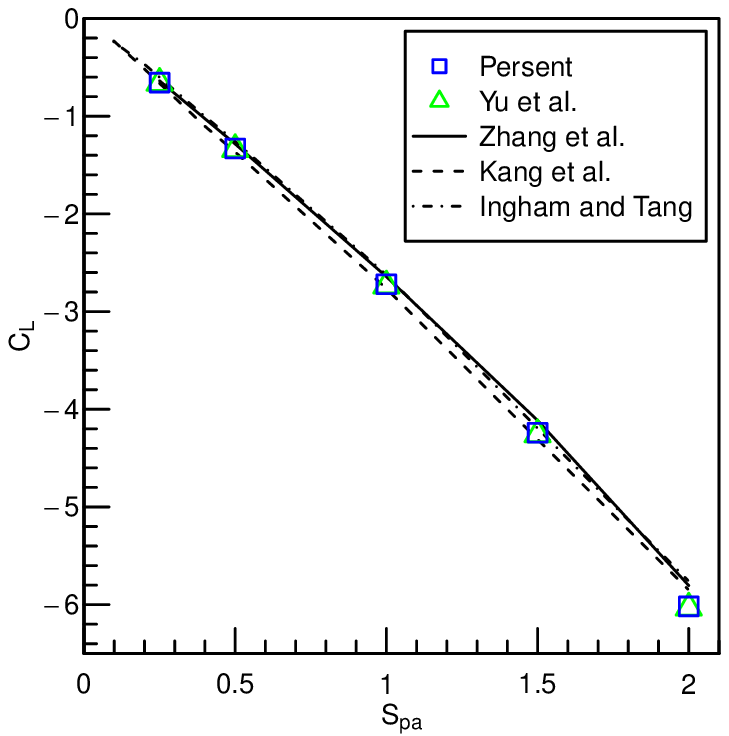}
\caption{Lift coefficient $C_L$ as a function of the spin number $S_{pa}$.}
\label{fig:spa_vs_cl}
\end{centering}
\end{figure}

To further validate the proposed scheme for more complex flows, simulations of a rotating particle immersed in a Poiseuille flow are also carried out. Non-equilibrium extrapolation scheme is used for the left velocity inlet and zero velocity gradient outlet for the right side. Zero velocity gradient is achieved by simply modify all distributions at the boundary equal to the distributions at closest fluid nodes. The top and bottom boundaries are set as solid walls. The particle spins with a constant angular velocity $\omega$. An additional lift force acts on the particle due to the rotation. This phenomenon is due to the well-known Magnus effect. To speed up the convergence of equilibrium state, the fluid field is initialized to the Poiseuille flow as $u_x= 4U\frac{y}{L}(1-\frac{y}{L})$, where $U$ is the unperturbed mainstream speed (along with the x-direction). A lift coefficient, which indicates the magnitude of the lift force, can be obtained as $C_L = \frac{F_y}{\rho_f U^2 R}$ depending on the lift force $F_y$. Here, the Reynolds number is defined as $Re=\frac{2 U R}{\nu}$. Another dimensionless number is the spin number $S_{pa}=\frac{\omega R}{U}$. The domain size is $L=400$ and radius of the particle $R=10$. Fluid properties are fixed as $U=0.05, \nu=0.02$. Fig.~\ref{fig:spa_vs_cl} shows the lift coefficient varying with the spin number at $Re=20$. It also shows an excellent agreement with the results of Kang et al.~\cite{kang1999laminar} and Ingham and Tang~\cite{ingham1990numerical}.

\section{Concluding remarks}\label{sec:conclusion}
A velocity interpolation based Bounce-Back scheme has been developed in this study. We combine the ideas of interpolated bounce-back schemes and non-equilibrium extrapolation scheme. The proposed scheme is validated by numbers of well-defined benchmark cases including Poiseuille flow with a moving wall, cylindrical Couette flow and Magnus effect. It is shown that the proposed scheme provides better accuracy at high Reynolds number and less dependency on solids positions which may crucial in many applications.

\section*{Acknowledgement}\label{sec:Acknowledgement}



%% References


\bibliographystyle{elsarticle-num}

\bibliography{mybib.bib}






% \end{thebibliography}

\end{document}